\documentclass[twocolumn,secnumarabic,amssymb,longbibliography, nobibnotes, aps, pra]{revtex4-2}

\usepackage{amsmath}
\usepackage[utf8]{inputenc}
\usepackage{xcolor}
\usepackage{bbold}
\usepackage{ulem}
\usepackage{hyperref}
\usepackage{newfloat}
\DeclareFloatingEnvironment[
fileext=loa,
listname={List of Algorithms},
name=Algorithm,
placement=tbp
]{algorithm}
\usepackage{algpseudocode}
\usepackage{comment}
\usepackage{natbib}
\usepackage{physics}
\usepackage{graphicx}
\hypersetup{
    colorlinks=true,
    linkcolor=red,
    citecolor=magenta,
    filecolor=magenta,
    urlcolor=cyan,
    runcolor=cyan
}
\begin{document}

\title{Optimizing Entanglement Distillation Policies via Markov Decision Process Formulation}

\author{Jigyen Bhavsar, Rajni Bala, Siddhartha Santra}
\affiliation{Department of Physics and Center of excellence in Quantum information, computation, science and technology, Indian Institute of Technology Bombay, Mumbai, Maharashtra 400076, India}

\begin{abstract}
    Entanglement distillation is a fundamental operation in quantum information processing used to obtain higher-fidelity entangled pairs from a supply of less entangled quantum states using local operations aided by classical communication (LOCC). In a physically relevant setting, where states with an initial fidelity of $f_0$, probabilistically generated over multiple, $m$, memory pairs distributed between two parties, Alice and Bob, are pairwise distilled, the optimal policy identifies the system-configuration dependent sequence of entanglement generation and distillation operations that need to be performed in order to minimize the expected time to reach some target fidelity $f_T>f_0$. Here, we formulate and systematically analyze this task as a Markov decision process and using a value iteration algorithm, obtain optimal deterministic policies that minimize the expected waiting time required to achieve a target fidelity. Our results show that the expected waiting time under the optimal policy decreases with increasing generation probability $p$ and number of quantum memories $m$ - as expected. In contrast, it exhibits non-monotonic behavior with respect to $f_0$ for a fixed fidelity gap, $(\Delta f = f_T-f_0)$. While the optimal policy consistently outperforms baseline policies such as the greedy, nested and entanglement pumping policies, its relative advantage is regime-dependent, being determined by the system parameters ($p,f_0,f_T,m$), and exhibits a nontrivial dependence on the fidelity gap $\Delta f$. Our results highlight the value of formulating entanglement distillation as a Markov decision problem, enabling the systematic design of policies that achieve target fidelity thresholds for quantum information tasks in realistic resource-constrained settings.

    %We compare the optimal policies with baseline policies such as the greedy, nested and entanglement pumping strategies. Our results show that the expected waiting time under the optimal policy decreases with increasing generation probability $p$ and number of quantum memories $m$, and exhibits non-monotonic behavior with respect to $f_0$ for a fixed fidelity gap, $(f_T-f_0)$. Moreover, the optimal policy outperforms baseline policies, with a regime-dependent advantage determined by the system parameters ($p,f_0,f_T,m$). Thus, our approach provides the minimum achievable expected time for reaching a desired fidelity threshold under optimal entanglement distillation policies in a physically relevant setting. 
\end{abstract}
\maketitle
\section{Introduction}

%Entanglement is a central resource for quantum information processing \cite{EPRPhysRev.47.777} whether it be communication, computation or metrology.  However, due to the stochasticity of the entanglement generation processes and the decoherence of qubits, quantum memories and quantum channels, producing high fidelity entangled pairs at sufficient rates is difficult in practice. To boost the quality of distributed entanglement between two entanglement distillation \cite{DurBriegel2016PurificationDistillation}. At present, entanglement distillation is particularly attractive for near-term quantum environment architectures, Entanglement distillation protocols convert multiple low-fidelity entangled pairs into a smaller number of high-fidelity pairs using only local operations and classical communication (LOCC). Such entanglement distillation protocols have been experimentally demonstrated across a variety of physical platforms \cite{Pan2001EntanglementPurificationNature,Pan2003,PhysRevLett.127.040506,PhysRevLett.128.080504} and constitute a promising pathway toward near-term realizations of measurement-based quantum computation \cite{PhysRevA.68.022312}, distributed quantum computation \cite{8910635,CuomoCaleffiCacciapuoti2020DistributedQCEcosystem}, distributed quantum sensing \cite{PhysRevLett.120.080501,Zhang_2021}, and quantum internet \cite{Kimble2008,Wehner2018}.\\

Entanglement is a central resource for quantum information processing \cite{EPRPhysRev.47.777}, enabling applications in communication \cite{Wehner2018}, computation \cite{8910635,CuomoCaleffiCacciapuoti2020DistributedQCEcosystem}, and metrology \cite{PhysRevLett.120.080501,Zhang_2021}. Nevertheless, producing high-fidelity entangled states at sufficient rates between spatially separated quantum systems remains a significant practical challenge due to the stochastic nature of entanglement generation processes and the effects of decoherence in the qubits or quantum memories as well as the quantum channels connecting them.

In principle, entanglement distillation operations \cite{Bennett1996Purification, Deutsch1996DEJMPS, DurBriegel2016PurificationDistillation} can mitigate these challenges by converting multiple low-fidelity entangled pairs into a smaller number of high-fidelity pairs using local operations and classical communication (LOCC) and simple variants of such operations have been experimentally demonstrated across a variety of physical platforms \cite{Pan2001EntanglementPurificationNature,Pan2003,PhysRevLett.127.040506,kbw2-fdqn,PhysRevLett.128.080504}. An important consideration in the design of a distillation policy, defined by the selection and ordering of entanglement generation and distillation operations for a given configuration of quantum memory pairs and connecting quantum channels, is the waiting time required to deliver a state with fidelity at or above a specified target value. Because quantum memories experience decoherence while stored states await further processing, longer waiting times can substantially reduce the quality of the resulting entangled state. Optimizing the distillation policy is therefore essential for minimizing waiting times and maximizing the delivery rate of high-fidelity entanglement.

While several useful policies such as entanglement pumping \cite{EntanglementPumping}, nested purification \cite{Dur1999Nestedpurification}, greedy distillation policy \cite{Ladd2006HybridQuantumRepeater} and banded purification \cite{4695947} work well in specific situations, the space of possible distillation policies is large, and finding the optimal policy can benefit from a formulation that casts it as a Markov decision process (MDP) \cite{SuttonBarto2018RL} especially under realistic resource and operational constraints. This is suggested by the fact that MDP formulations have been demonstrated to be useful in boosting the performance of quantum protocols in different contexts -- with the common feature of sequential decision-making - ranging from the design of optimal pulse sequences and control strategies in noisy environments \cite{PhysRevX.8.031086,Niu2019}, for optimizing and fault-tolerantly adapting quantum error correction schemes across different noise models \cite{Nautrup2019OptimizingQEC} and for determining optimal entanglement swapping protocols for entanglement distribution \cite{PhysRevLett.128.150502,Inesta2023OptimalPolicies} and optimal routing strategies in quantum networks \cite{11275902}.

In the context of entanglement distillation, we consider the following concrete scenario. Two parties, $A$ and $B$, share $m \geq 2$ independent and identical quantum channels connecting $m$ pairs of long-lived quantum memories in their possession, and have the ability to implement two types of operations - entanglement generation (EG) and 2-to-1 entanglement distillation (ED) - which they seek to use in a system configuration-dependent sequence of steps to obtain a shared entangled state with fidelity at least equal to a desired target, while minimizing the expected time required to do so, see Fig. (\ref{MDP Env}). In our model, the primary resource constraint is the number of quantum memories $m$ at each node. Entanglement generation (EG) and 2-to-1 entanglement distillation (ED) operations can be performed in parallel on disjoint sets of memory pairs, with the constraint that no memory participates in more than one operation at a time. In particular, EG can be applied simultaneously on all inactive links, while multiple ED operations may be executed in parallel on stored entangled pairs. As a special case, one may restrict the system to allow at most a single distillation operation per step, corresponding to more constrained hardware settings where parallel two-qubit operations are limited by crosstalk and residual couplings, leading to correlated errors and degraded fidelity \cite{Sarovar2020,Heinz2021,Cheng2024}.

Both entanglement generation (EG) and entanglement distillation (ED) are inherently probabilistic processes. While an EG attempt over any of the $m$ channels succeeds with a fixed probability $p$ and produces a state with fidelity $f_0$, the success probability of an ED attempt depends on the fidelities of the two input states used in the distillation operation. Moreover, EG and ED operations require different execution times, which depend on the specific protocols employed \cite{Duan2001,Moehring2007,Zukowski1993,Pan1998,Cabrillo1999,Barrett2005,Munro2015}, primarily due to the classical communication needed to herald their success or failure (neglecting the time required for local operations).

At any stage of the sequence of operations determined by a distillation policy, the configuration of the $m$ memory pairs can be described as a set of unentangled pairs, raw entangled pairs (of fidelity $f_0$), and pairs in intermediate states ( of fidelities $f_0\leq f<f_T$) resulting from prior successful distillation steps. Given that information about the system configuration is accessible to Alice and Bob, they have the freedom to decide which pair, if any, to use in a 2-to-1 ED operation, and which links, if any, to attempt EG on, thereby leading to the next (probabilistically realised) configuration of the $m$ quantum memory pairs. Because the configuration-dependent choice of actions at each step determines the subsequent trajectory through the configuration space, different sequences of actions (i.e., different distillation policies) yield different probabilities—and hence different expected times—for reaching a desired target fidelity $f_T$. The optimal policy we seek is a map from the space of system configurations to the space of possible actions that minimizes the expected time to obtain a single state with fidelity at least equal to $f_T$.

In this paper we investigate this sequential decision-making problem formulated as a finite-state MDP problem, whose solution yields an optimal sequence of system configuration dependent actions—i.e., a distillation policy specifying both the operations and the memories involved—that, in general, depends on the system parameters $m$, $p$, $f_0$, and $f_T$, and minimizes the expected time required to obtain a single entangled state with fidelity $f \geq f_T$, thereby maximizing the rate of entanglement delivery at the target fidelity. In this formulation, the states of the MDP capture the fidelities of all stored elementary links and the actions correspond to either entanglement generation attempts \cite{Duan2001,Moehring2007,Munro2015} or selective 2-to-1 entanglement distillation operations \cite{Bennett1996Purification, Deutsch1996DEJMPS, DurBriegel2016PurificationDistillation}. We solve the resulting MDP using the value iteration algorithm \cite{SuttonBarto2018RL}, obtaining optimal deterministic entanglement distillation policies along with their corresponding minimum expected waiting times, and benchmark these policies against standard baselines such as greedy, nested and pumping policies. Our results show that the expected waiting time under the optimal policy decreases with increasing entanglement generation probability $p$ and the number of quantum memories $m$. More interestingly, for a fixed fidelity gap, $\Delta f$, between the initial and target fidelity, the optimal expected waiting time exhibits non-monotonic behavior with respect to $f_0$. Even under ideal operations and perfect quantum memories, the optimal policy provides substantial reductions in expected waiting time compared with baseline policies, with improvements reaching approximately 50$\%$ over the greedy policy in some parameter regimes and up to 80$\%$ over nested policy in others, depending on the system parameters $(p,f_0,f_T,m)$. However, the performance advantage of the optimal policy evolves nontrivially with  $\Delta f$, depending on the chosen baseline policy and the system parameters $(p,m,f_0)$.

The remainder of this paper is organized as follows. In Sec. \ref{system model and problem formulation}, we introduce the system model and formalize the entanglement distillation policy optimization problem. In Sec. \ref{MDP formulation}, we present its formulation as Markov decision process (MDP). In Sec. \ref{baseline policies}, we define the baseline ED policies used for benchmarking and comparison. In Sec. \ref{results}, we present numerical results for optimal policies and expected waiting times, including comparisons with baseline strategies. Finally, Sec. \ref{conclusions} concludes the paper and outlines future directions.

\section{System Model and Problem Formulation}
\label{system model and problem formulation}
\renewcommand{\thesubsection}{\Alph{subsection}}

In this section, we describe the physical model underlying entanglement generation and distillation. We then formalize the problem of entanglement distillation policy optimization within this framework.

%%%%%%%%%%%%%%%%%%%%%%%%%%%%%%%%%%%%%
\subsection{Model of entanglement generation and distillation }
\label{Entanglement distillation Model}

In our setup, two distant parties, Alice and Bob, each equipped with $m\geq 2$ quantum memories are connected by $m$ parallel quantum channels over each of which they can attempt probabilistic and heralded entanglement generation \cite{Sangouard2011RepeatersReview,Simon2007PhotonPairRepeaters,Sangouard2009EncodedRepeaterA79,Knill2005A72}  independently. An entanglement generation attempt over a single channel succeeds with probability $p$ resulting in a Werner state,
\begin{align}
    \rho(f)= f\,\ket{\psi^-}\!\bra{\psi^-}+\frac{(1-f)}{3}\sum_{\beta\in\{\psi^+,\phi^-,\phi^+\}}\ket{\beta}\!\bra{\beta}
    \label{rho_f}
\end{align}
of {\it raw} fidelity, $f=f_0$, and consumes some amount of time for the classical communication required for heralding. The raw fidelity is assumed to be above the threshold required for successful distillation, $1/2< f_0\leq 1$. Further, parallel entanglement generation operations over the channels are allowed in any step of the operational sequence.

%Entanglement generation is performed using a midpoint-heralding architecture, where each node locally prepares a photon–matter entangled state and transmits the photonic mode through the channel to the central station. The two photons interfere at a beam splitter and a suitable Bell-state measurement (BSM) pattern projects the two remote memories into an entangled state. This process establishes a bipartite quantum entangled states in a probabilistic and heralded manner, implementable using a variety of physical systems . Assuming the entangled pair sources generate maximally-entangled two-qubit pure states, the so called Bell states which are, $\{\ket{\psi^-},\ket{\psi^+},\ket{\phi^-},\ket{\phi^+}\}$, which suffer depolarazing noise in the quantum channels, the output of this process yields a state shared between neighboring nodes which takes the form,

Alice and Bob are also allowed to utilise, as needed, a 2-to-1 entanglement distillation operation utilising the Deutsch's protocol - which is also heralded and probabilistic \cite{Deutsch1996DEJMPS}. It utilises two states of the form Eq. (\ref{rho_f}), not necessarily of the same fidelity, and processes them jointly via local operations and measurements at the locations of Alice and Bob. Based on agreement of measurement outcomes communicated classically, they can obtain a single pair of higher fidelity with a probability that is function of the fidelities of the two states input to the ED process.

When the distillation operation succeeds, the output of a 2-to-1 ED operation that utilises two input Werner states of fidelities, $f_1,f_2$ is another Werner state of fidelity,
%{\color{red} Bell diagonal state}
\begin{align}
    f_d &= \frac{f_1f_2 + \frac{(1-f_1)(1-f_2)}{9}}{f_1f_2 + \frac{f_1(1-f_2) + f_2(1-f_1)}{3} + \frac{5}{9}(1-f_1)(1-f_2)},
    \label{fd}
\end{align}
which occurs with a probability,
\begin{align}
    p_d &= f_1f_2 + \frac{f_1(1-f_2) + f_2(1-f_1)}{3} + \frac{5}{9}(1-f_1)(1-f_2),
\label{pd}
\end{align}
while consuming some amount of time for the necessary classical heralding signals.

Note that the amounts of classical communication time needed for the two operations, EG and ED, can be different and the relative magnitudes of these times determine the policy that is found optimal under the value iteration algorithm.

%The ED process requires classical communication overhead to convey the measurement outcomes. This heralding delay directly contributes to the overall waiting time in entanglement distribution, as describe in Sec. \ref{MDP formulation} in detail. With multiple memories, the experimentalist must decide which pairs to distill and when, which balances the quick fidelity boost against using valuable entanglement pairs and waiting for classical communication delays. This naturally leads to a scheduling problem, which we formalize next as an optimal control task.

%the parties begin by applying fixed local $\pi/2$ basis rotations that symmetrize the noise and map the shared states into a form optimized for distillation. Two imperfect pairs are then processed jointly: one is designated as the source and the other as the target. Each party performs a CNOT from the source qubit to the target qubit and subsequently measures the target qubit in the computational basis. After exchanging their measurement outcomes, the parties retain the source pair only when the results coincide. 

%%%%%%%%%%%%%%%%%%%%%%%%%%%%%%%%%%%%%
\subsection{Entanglement distillation policies}
\label{Entanglement distillation Scheduling}

For a given target fidelity, $f_T$, a single round of entanglement distillation may generally be insufficient to produce an entangled pair with fidelity $f\geq f_T$. Instead, multiple distillation steps must be performed, where the output pair from one successful step is retained and subsequently used as an input to later steps. This produces a sequence of distillation steps in which entanglement is progressively upgraded at the expense of consuming additional entangled pairs. With multiple quantum memories at each party, some memories store entangled pairs that may be selected for distillation while others are used for entanglement generation. The experimentalist must decide whether and which stored entangled pairs should be distilled or preserved for future use. Since both entanglement generation and distillation are probabilistic and time-consuming, these decisions strongly influence the overall waiting time to reach the target fidelity. By choosing the optimal entanglement distillation policy, one seeks to obtain a target-quality entangled pair as quickly as possible for a given number of quantum memories and operational constraints.

At any stage in the sequence of operations, Alice and Bob share a configuration of stored entangled pairs across their $m$ quantum memories. This configuration is specified by identifying which memory pairs are empty ($f=0$) and by the fidelities of the stored entangled pairs ($f_0 \leq f \leq f_T$). Depending on this configuration, they can either attempt entanglement generation on selected empty memory pairs or perform 2-to-1 ED operation in parallel on a chosen pair of stored entangled states. The choice of action at each step determines how the system evolves over time and which configurations are subsequently realized before reaching the target fidelity. An entanglement distillation policy, denoted by $\pi$, is thus a rule that specifies which action to take based on the current configuration of the system. The problem can thus be stated as follows: \textit{given the current set of stored entangled pairs and their fidelities, which action should be performed next, in order to minimize the expected time required to obtain at least one entangled pair with fidelity  greater than or equal to target threshold $f_T$?}

%The policy operates in discrete decision epochs. At each round, the experimentalist has two possible choices. First, Alice and Bob may attempt to generate entanglement using any currently unused memory pairs.Each attempt succeeds with probability $p$, and upon success the resulting shared state is a Werner state $\rho$ with initial fidelity $f_0$, as given by eq. \eqref{rho_f}. Second, whenever at least two entangled pairs are simultaneously available in memory, Alice and Bob may choose to apply a 2-to-1 entanglement distillation step. This step consumes two stored pairs with fidelities $(f_1,f_2)$, succeeds with probability $p_d(f_1,f_2)$, as given by eq. \eqref{pd}, and produces a new pair of improved fidelity $f_d(f_1,f_2)$, as given by eq. \eqref{fd} when successful, while discarding both pairs upon failure. Since both entanglement generation and distillation are probabilistic and incur classical communication delays, the ordering of these actions strongly influences the overall waiting time. The scheduling problem can thus be stated as follows: \textit{given the current set of stored entangled pairs and their fidelities, which action should be performed next, in order to minimize the expected time required to obtain at least one entangled pair with fidelity above a target threshold $f_T$?}\\

To formalize this objective, we define the waiting time $t_{\pi}$ as a random variable representing the time required for the system, under a given ED policy $\pi$, to reach an absorbing state in a single realization of the stochastic process. The performance metric of interest is the expected waiting time,
\begin{align}
T_{\pi} = \mathbb{E}[t_{\pi}]
\end{align}
where the expectation is taken over the stochastic outcomes of entanglement generation and distillation conditioned on the decisions prescribed by policy $\pi$. Our goal is to find an optimal policy that minimizes the expected waiting time, defined as follows, 
\begin{align}
\pi^{*} = \arg\min_{\pi}T_{\pi},
\end{align}
and the corresponding minimum expected waiting time, $T_{\pi^*}$. This optimization problem is a sequential decision-making task with stochastic transitions, which naturally motivates the use of a Markov decision process (MDP) framework to compute the optimal ED policy, as described in the next section.

%%%%%%%%%%%%%%%%%%%%%%%%%%%%%%%%%%%%%%%%%%%%%%%%%%%%%%%%%%%%%%%%%%%%%%%%%%
\section{Markov Decision Process Formulation}
\label{MDP formulation}

\label{MDP formulation}
\begin{figure}
    \centering
    \includegraphics[width = \linewidth]{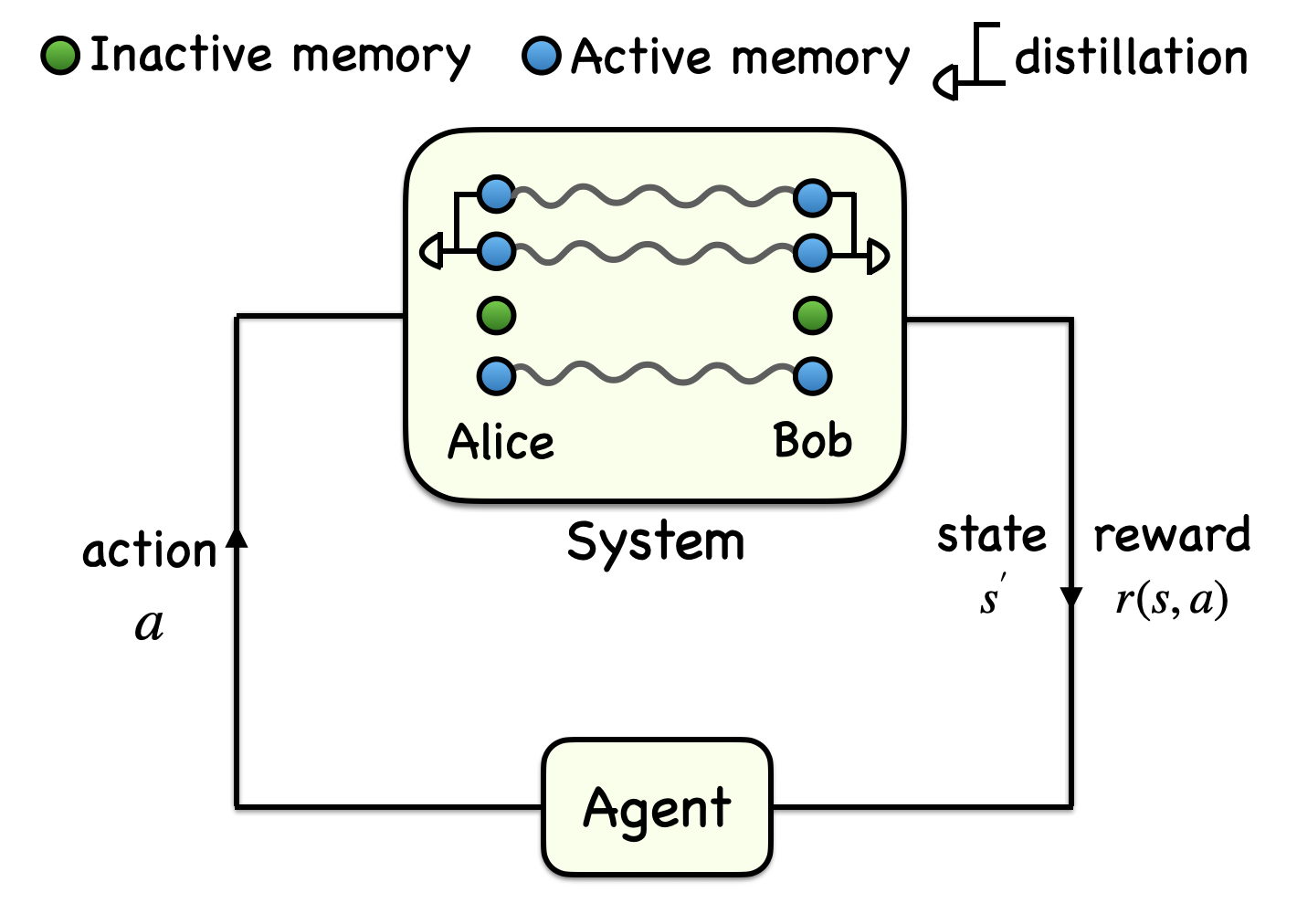}
    \caption{Schematic representation of MDP formulation for optimizing entanglement distillation policy. The agent interacts with system, which consists of two distant  parties, Alice and Bob, each equipped with $m\geq 2$ quantum memories. In above configuration, three quantum memories are active (blue circle) that stores the entangled pairs (curved gray lines), while the remaining memories are inactive (green circle). The pairs in the first and second active memories are selected for distillation , while the other pairs remain unchanged. At each decision step, the agent selects an action $a$ (entanglement generation or distillation), after which the system transitions to a new state $s^{'}$ from the current state $s$, representing the updated configuration of stored entangled pairs, and returns a reward $r(s,a)$ corresponding to the incurred time cost. }
    \label{MDP Env}
\end{figure}

%We model the search for an optimal entanglement distillation policy as a Markov decision problem, which provides a principle framework for policy optimization, for which many methods have been developed, such as dynamic programming methods, Q-learning and deep reinforcement learning \cite{SuttonBarto2018RL}. 

We model the problem of optimizing entanglement distillation policies as a Markov decision process (MDP)\cite{SuttonBarto2018RL}, which provides a systematic framework for sequential decision-making under uncertainty. Formally, an MDP is defined by the 4-tuple $(\mathcal{S},\mathcal{A},\mathcal{P},\mathcal{R})$. Here, $\mathcal{S}$ is the set of all possible states $s\in \mathcal{S}$ that the agent can observe, $\mathcal{A}$ is the set of all possible actions $a\in \mathcal{A}$ that the agent can take, $P(s^{'}|s,a)$ is the transition probability of system (referred to as the “environment” in MDP literature) moving from state $s\in \mathcal{S}$ to $s^{'}\in \mathcal{S}$ under action $a\in \mathcal{A}$, and $\mathcal{R}$ is the values for the reward function $r(s,a) \in \mathcal{R}$. In MDP, at each time step the agent selects an action $a$ based on the current state $s$ of the system. The system then transitions to a new state $s^{'}$, according to the transition probability $P(s^{'}|s,a)$ and returns a reward $r(s,a)$ to the agent, as shown in Fig. (\ref{MDP Env}). The agent’s objective is to learn a policy $\pi(a|s)$, i.e. mapping from state to action, that maximizes the expected cumulative reward.

We consider an agent located at the midpoint of the parallel quantum channels connecting Alice and Bob. At each time step, the agent observes the state of the system and selects an action according to the policy $\pi$. The agent has two types of control actions i.e. entanglement generation and 2-to-1 entanglement distillation. For entanglement generation, the agent identifies unused quantum memories at each node and sends a request to Alice and Bob to create new elementary entangled pairs, which incurs a time delay, referred to as the request time. We adopt a midpoint-heralded architecture \cite{Duan2001,Moehring2007,Munro2015} in which the success of entanglement generation is determined at the midpoint. Assuming that the creation of local photon–matter entangled states and the Bell-state measurement (BSM) at the midpoint are instantaneous, the total classical communication (CC) time between initiating the entanglement generation action and being able to take the next action is $l/c$, where $l$ is the link distance and $c$ is the speed of light. This delay accounts for both the request signaling and the photonic transmission to the midpoint station where the BSM is performed. Once the probabilistic BSM outcome is obtained, the agent updates its knowledge of the system and proceeds to the next decision.

For entanglement distillation, the agent monitors the set of available entangled pairs shared between Alice and Bob. If at least two entangled pairs are available, the agent may send a request for Alice and Bob to perform a 2-to-1 ED using Deutsch protocol \cite{Deutsch1996DEJMPS} on a selected pair of stored links. As discussed in Sec. \ref{Entanglement distillation Model}, distillation requires exchanging measurement outcomes between the end nodes, which introduces an additional CC overhead. Consequently, the total time delay associated with a distillation action is $2l/c$, incorporating the request time, the time required for outcome exchange, and the time needed to report the result back to the agent after the local operations are completed. Moreover, a single entanglement distillation operation consumes twice as many quantum resources compared to entanglement generation. These action-dependent time and resource costs make the entanglement distillation policy optimization problem inherently nontrivial.

In a single decision step, we permit parallel entanglement generation across all unused quantum memories, as well as multiple 2-to-1 entanglement distillation operations on disjoint pairs of stored links. Each memory participates in at most one operation per step. As a special case, one may restrict the system to allow only a single distillation operation per step, reflecting more constrained hardware settings where parallel two-qubit operations are limited by crosstalk and correlated errors \cite{Sarovar2020,Heinz2021,Cheng2024}. The construction of the MDP used in this work is described as follows.

%In a single decision step, we explicitly permit the generation of parallel entanglement across several unused quantum memories. However, we only permit atmost one entanglement distillation operation per step, as parallel two-qubit operations are prone to crosstalk and correlated errors \cite{Sarovar2020,Heinz2021,Cheng2024}. The construction of the MDP used in this work is now summarized as follows. 
%The detail description is provided in the Appendix \ref{MDP Formulation}. 

\noindent\textit{(a) State space:} The state of the system represents the configuration of fidelities of entangled states stored across the quantum channels connecting Alice and Bob. Given an initial fidelity $f_0$ and target fidelity $f_T$, there exists a discrete set of attainable fidelity levels arising from the recursive structure of the 2-to-1 entanglement distillation, which progressively generates higher-fidelity states from the initial pairs. Formally, the state space is given by $S \in \{0,f_0,...,F_T\}^m$, where each component specifies the fidelity of a corresponding link, taking values from this set of allowed fidelity levels. Any fidelity greater than or equal to the target fidelity, i.e., $f \geq f_T$, is represented by a single absorbing state $F_T$ in the state space of MDP. At any time step $t$, the state of the system is represented as, 

\begin{align}
    \vec{s}_t = \{f_1,f_2,...,f_m\} \in \mathcal{S}
\end{align}
where, $m$ is the total number of quantum memories at each node and element $f_j$ of the state $\vec{s_t}$ is the random variable which gives the fidelity of entangled link stored in $j^{th}$ memory from the allowed fidelity level. The $f_j = 0$ indicates that the $j^{th}$ parallel elementary link is inactive and $f_j \neq 0$ indicates that the $j^{th}$ parallel elementary link is active. Consequently, a finite state space exists only for target fidelities satisfying $f_T < f_{\infty}(f_0)$ \cite{RSS}. Here, $f_{\infty}(f_0)$ denotes the asymptotic fidelity attainable through repeated successful pumping with entangled pairs of initial fidelity $f_0$, given by,

\begin{align}
    f_{\infty}(f_0) = \frac{-3 + 6f_0 + \sqrt{7-26f_0 + 28f_0^2}}{-2+8f_0}
    \label{finf}
\end{align}
 
In order to overcome that limitation, we developed the approximate method by dividing the fidelity gap between initial fidelity and target fidelity in small bins and each bin represents the state in the MDP. However, this method becomes inefficient for large number of memories, $m$, and fidelity gap $\Delta f$.\\
 
 %resulting from entanglement pumping 
\noindent\textit{(b) Action space:} Due to heralding, the state is assumed to be fully known to the agent at all time. At each decision step, the agent selects an action based on the current state of the system. The available actions depend on the state, can be broadly classified into two categories:

\textit{(i) Entanglement generation ($\mathcal{A}_g$) :} This action specifies, for each elementary link, whether an entanglement generation attempt is made. It is represented as, 
\begin{align}
    a_g = (a_g^1, a_g^2,...,a_g^m) ; a_g \in \{0,1\}^m
\end{align}
where, $a_g^i = 1$ indicates that an entanglement generation attempt is performed on the $i^{th}$ link and $a_g^i = 0$ indicates that no generation attempt is made and the current state of the link is retained.

\textit{(ii) Entanglement distillation ($\mathcal{A}_d$) :} This action allows simultaneous 2-to-1 distillation across multiple pairs of parallel links within single time step. A distillation action is specified by a matching $M$ on the set of available links $\{1,2,...,m\}$, defined as a set of disjoint pairs, 
\begin{align}
    M = \{(i_1,j_1),(i_2,j_2),...,(i_k,j_k)\}, \ j_l>i_l, \ \forall l
\end{align}
where all indices are distinct, i.e., no link participates in more than one distillation pair simultaneously. The set of all valid matchings, including the empty matching $M = \emptyset$ (no distillation), is denoted $\mathcal{M}$.

The full action at each time step is represented by a tuple $a = (M,a_g)$ specifying distillation matching and generation vector on the remaining free links. The combined action space is,
\begin{align}
    \mathcal{A} = \bigcup_{M\in \mathcal{M}} \{(M,a_g) ; a_g \in \{0,1\}^{m-2|M|} \}
\end{align}
The total number of actions is given by,
\begin{align}
    |\mathcal{A}| = \sum_{k=0}^{\lfloor m/2 \rfloor} \frac{m!}{(m-2k)! k! 2^k}.2^{m-2k}
\end{align}
where the first factor counts the number of distinct matchings of size $|M| = k$ over $m$ links, and the second factor enumerates generation choices on the $m-2k$ free links. The overall action space size scales as $O(m^{\lfloor m/2 \rfloor}.2^m)$.

This construction generalizes the special case in which at most one 2-to-1 ED is allowed per step ($|M| \leq 1$). In this setting, $(M = \emptyset, a_g)$ corresponds to pure generation across all $m$ links, and ($M = \{(i,j)\},a_g$) corresponds to a single distillation between links i and j with generation on the remaining $m - 2$ links. By permitting arbitrary matchings $M\in \mathcal{M}$, the extended action space enables fully parallel, state-dependent distillation across multiple link pairs simultaneously, while maintaining a finite action space.\\

\noindent \textit{(c) Reward function:}  The objective of the problem is to minimize the total expected waiting time required to obtain at least one entangled pair whose fidelity exceeds a prescribed target fidelity, $f_T$. In order to encode this goal within the MDP framework, the rewards are selected so that, when the reward is maximized, the waiting time resulting from classical communication is minimized. In the units of one-way classical communication delay($l/c$), the reward function evaluates to $r = -1$ for entanglement generation and $r = -2$ for entanglement distillation. When the terminal state is reached, the reward function evaluates to $r = 0$. This choice enforces minimization of the expected waiting time.\\

\noindent \textit{(d) System dynamics:} As entanglement generation and entanglement distillation succeed only probabilistically, therefore, the dynamics of the system are stochastic and are described by the transition probabilities $P(s'|s, a)$. Entanglement generation succeeds independently on each unused memory pair with a fixed probability $p$, while distillation succeeds probabilistically according to the fidelities of the two input pairs as given by Eq. (\ref{pd}) and, upon success, the fidelity of the new state is given by Eq. (\ref{fd}).\\

As defined in Sec. \textcolor{red}{II}\ref{Entanglement distillation Scheduling}, $T_{\pi}(s)$ denotes the expected waiting time, i.e., the expected total time required to reach a target state when starting from state $s$ and following a policy $\pi$. The expected waiting time $T_{\pi}(s)$ satisfies the Bellman equation,
\begin{align}
    T_{\pi}(s) = r\big(s,\pi(s)\big) + \gamma \sum_{s' \in \mathcal{S}} P\big(s' \mid s, \pi(s)\big)\, T_{\pi}(s'), \quad \forall s \in \mathcal{S},
    \label{bellman eq.}
\end{align}
where $\mathcal{S}$ denotes the state space, $\gamma$ is the discount factor, $P(s' \mid s, a)$ is the transition probability from state $s$ to state $s'$ under action $a$, and $r(s,a)$ represents the time cost associated with taking action $a$ in state $s$.

An optimal policy can be found by minimizing $T_{\pi}(s)$ $\forall s\in \mathcal{S}$, using Eq. (\ref{bellman eq.}). To solve this optimization problem, we employ a dynamic programming approach based on the value iteration algorithm, as described in Appendix \ref{Dynamic programming algo} and for more detail, see chapter 4 from \cite{SuttonBarto2018RL}.  Fig. (\ref{distillation figures}\textcolor{red}{d}), shows the one possible history of the optimal policy for the parameter regime $p = 0.95, f_0 = 0.648$ and $f_T = 0.710 $.

\begin{comment}
In order to compute the optimal scheduling policy for entanglement distillation, we adopt a dynamic programming approach and solve the corresponding MDP using the value iteration algorithm. Due to fully specified environmental dynamics, the value iteration algorithm is well suited to our limited setting with a finite state space and a finite action space, as describe in detail in Appendix \ref{Dynamic programming algo}. The state space size grows rapidly with the number of memories $m$ and the number of fidelity levels $N(f_0,f_t)$, which depends on the initial and target fidelity. In particular, the state space scales as  $|S| = (N(f_0,f_t) + 2)^m$, leading to exponential growth in computational complexity. This exponential scaling restricts exact dynamic programming to environment with a finite number of fidelity levels. For larger memory sizes or additional degrees of freedom like memory decoherence, approximate methods like reinforcement learning algorithms become more suitable.

Due to fully specified environmental dynamics, the value iteration algorithm can be used for our limited setting with a finite state space and a finite action space
\end{comment}

%%%%%%%%%%%%%%%%%%%%%%%%%%%%%%%%%%%%%
%%%%%%%%%%%%%%%%%%%%%%%%%%%%%%%%%%%%%
\section{Baseline Policies}
\label{baseline policies}

\begin{figure}[t]
    \centering
    \includegraphics[width = \linewidth]{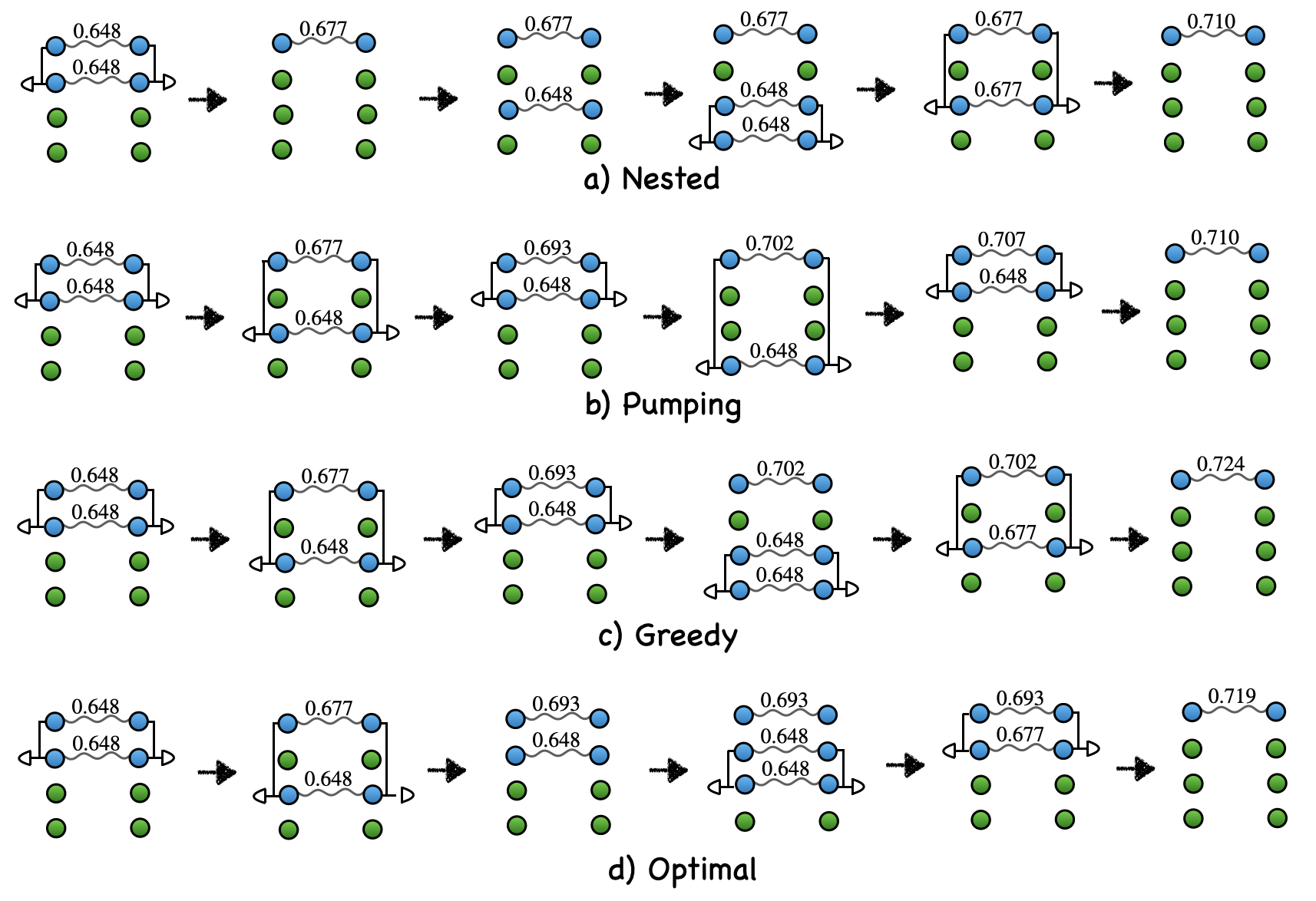}
    \caption{Time evolution of the history trees for different entanglement distillation policies with four quantum memories at each station. The initial fidelity after entanglement generation (with probability $p = 0.9$) is 0.648, and the target fidelity is 0.710. The grey curved lines denote entangled pairs, with the associated numbers indicating their fidelities. Blue circles represent active quantum memories, while green circles indicate inactive quantum memories. (a) History tree of entangled pairs under the nested purification policy. (b) History tree of entangled pairs under the entanglement pumping policy. (c) History tree of entangled pairs under the greedy policy.  (d) History evolution for an example optimal policy obtained from our MDP formulation.  }
    \label{distillation figures}
\end{figure}

A variety of heuristic entanglement distillation policies have been proposed in the literature \cite{EntanglementPumping,Dur1999Nestedpurification,Ladd2006HybridQuantumRepeater,4695947}. In certain limiting cases, the optimal policy may approach some of these known strategies. These baseline policies are attractive because they are simple to implement, such as “distill whenever possible” or “only distill pairs of similar quality”. The policies like entanglement pumping, nested purification and greedy policy, performs well only within specific parameter regimes ($m,p,f_0,f_T$) of the system. For this reason, we use them as benchmarks to evaluate and highlight the performance gains achievable with the optimal policy. We now summarize the baseline policies considered in this work.

\noindent \textit{(a) Pumping policy}: This scheme was originally proposed by Dür \textit{et al.}\cite{EntanglementPumping} that requires a minimum two quantum memory per node, which can be extended to the multi-memory setting, as shown in Fig. (\ref{distillation figures}\textcolor{red}{b}) and expressed as stationary policy $\pi_{\text{pump}} : \mathcal{S} \rightarrow \mathcal{A}$ within the proposed MDP framework. At each decision step $t$, given the state $\vec{s}_t = \{f_1, f_2, \dots, f_m\}$, the policy prioritizes the use of elementary pairs of initial fidelity $f_0$ as a resource to “pump” higher-fidelity links. The action selection proceeds according to the following rules:

\textit{Entanglement generation:} 
For all inactive memories ($f_j = 0$), the policy sets $a_g^j = 1$ attempt generation, while leaving all occupied memories unchanged ($a_g^j = 0$ for $f_j \neq  0$)

\textit{Entanglement distillation:
Let $\mathcal{L}_0 = \{j:f_j = f_0\}$} be the set of links at initial fidelity. A distillation matching $M$ is constructed greedily that is, each link $j \in \mathcal{L}_0$ is paired with the unmatched active link of maximum fidelity among all remaining unmatched links. If no higher-fidelity partner ($f>f_0$) is available, $f_0$ links are paired among themselves. All resulting pairs are distilled simultaneously in a single time step.

This policy preserves the key intuition of entanglement pumping i.e., low-fidelity elementary pairs are repeatedly used to incrementally enhance a higher-fidelity link. The pumping policy is memory-efficient, but it improves the fidelity of entanglement very slowly. In particular, when the gap between the initial fidelity and the target fidelity is large, pumping becomes increasingly ineffective.\\

\noindent \textit{(b) Nested policy} :  This scheme, proposed by Dür \textit{et al.} \cite{Dur1999Nestedpurification}, in which the selected action is to perform distillation only between pairs of identical fidelity in state configuration of $m$ quantum memory pairs, as illustrated in Fig. (\ref{distillation figures}\textcolor{red}{a}). At each decision step $t$, given the system state $\vec{s}_t = \{f_1, f_2, \dots, f_m\}$, the policy $\pi_{nested}$ operates as follows :

\textit{Entanglement generation:}
Generation is attempted on all inactive links simultaneously, i.e., $a_g^j = 1 \ \forall j$ with $f_j = 0$, in parallel with any allowed distillation actions on the remaining links.

\textit{Entanglement distillation:}
Let $\mathcal{F} = \{f:|\{j: f_j = f\}|\geq 2\}$ be the set of fidelity levels with at least two active links. For each $f \in \mathcal{F}$ , links in $\mathcal{L}_f = \{j:f_j = f\}$ are paired into $\lfloor |\mathcal{L}_f|/2\rfloor$ disjoint pairs. The distillation matching is constructed as,
\begin{align}
    M = \bigcup_{f \in \mathcal{F}} \{(i,j):i,j\in \mathcal{L}_f,j>i,\text{disjoint}\}
\end{align}
If $|\mathcal{L}_f|$ is odd, one link at that fidelity remains idle until a compatible partner becomes available.

This policy enforces strict level-by-level distillation, combining pairs only of equal fidelity. However, exact fidelity matching requires long waiting times. In practical scenarios, when memory decoherence is present, the stored pairs shift to different fidelities over time, which makes the nested policy unrealistic. When resources are limited, this policy may even deadlock if no compatible pairs are available for distillation.\\

\noindent \textit{(c) Greedy policy}: The greedy strategy for entanglement distillation was introduced in \cite{Ladd2006HybridQuantumRepeater}. For any state configuration, containing at least two active links, the policy prioritizes immediate 2-to-1 ED action between the most compatible available pairs, i.e., those with the smallest fidelity difference. Fig. (\ref{distillation figures}\textcolor{red}{c}) shows one possible history tree of greedy strategy. Given the system state $\vec{s}_t = \{f_1, f_2, \dots, f_m\}$, the policy $\pi_{greedy}$ proceeds as follows:

\textit{Entanglement generation:} Generation is attempted on all inactive links simultaneously, i.e., $a_g^j = 1 \ \forall j$ with $f_j = 0$, as done in previous baseline policies.

\textit{Entanglement distillation:} Let $\mathcal{L}_{act} = \{j:f_j>0\}$ be the set of active links. If $|\mathcal{L}_{act}| >2$ a distillation matching $M$ is constructed greedily by iteratively selecting, among all remaining unmatched active link pairs, the pair $(i^*,j^*)$ with the smallest fidelity difference,
\begin{align}
    (i^*,j^*) = \arg \min_{\substack{i,j \in \mathcal{L}_{act} \\ \text{unmatched},j>i}}|f_i - f_j|
\end{align} 
and adding it to $M$ until fewer than two unmatched active links remain, yielding $ k = \lfloor |\mathcal{L}_{act}|/2\rfloor$ pairs. If $\mathcal{L}_{act}$ is odd, one active link remains idle. The full action is $a_t = (M,a_g) \in \mathcal{A}$, where $M$ is the minimum-$\Delta f$ greedy matching and $a_g$ applies generation on all inactive free links.

This policy aggressively exploits all available resources by pairing all active links simultaneously in increasing order of fidelity difference, maximizing resource utilization within each time step. Despite this, when fidelity differences are large, some pairings may still yield low distillation success probabilities and limited fidelity gain per step.

The relevance of these baseline strategies can be understood more clearly in certain limiting regimes of the system parameters ($m,p,f_0,f_T$). With no constraints on the availability of physical resources, $m\to\infty$, i.e., quantum memories per node, a pair of entangled states of fidelity $f_0$ is always available at any distillation step for generation probability $p>0$, then for any target fidelity $f_T\geq f_0$, the optimal policy converges to the well-known nested purification protocol \cite{Dur1999Nestedpurification}.
In contrast, when only two quantum memories ($m=2$) are available, the optimal policy reduces to the entanglement pumping protocol \cite{EntanglementPumping}, independent of probability of successful entanglement generation. In the intermediate regime of finite number of quantum memories and nonzero generation probability, our MDP-based framework determines the corresponding optimal policies, as shown in the next section.

\section{Numerical Results}
\label{results}
\renewcommand{\thesubsection}{\Alph{subsection}}

% para1 - explain general setup, problem statement, optimal policy, performance metric

As discussed in previous sections, the agent may attempt entanglement generation whenever no elementary entangled links are available between Alice and Bob. Once the entangled links have been established, the agent may perform distillation immediately in specific order or wait for later time. This sequence of decisions is determined by an entanglement distillation policy, i.e., a rule that maps system configurations to actions. An optimal policy is one that minimizes the expected waiting time to reach the target state, given an initial configuration with no entangled links. Throughout this work, we use the expected waiting time as the primary performance metric. We model the evolution of our quantum system as a markov decision process (MDP). Based on this formulation, we use the corresponding Bellman equation (see Eq. \ref{bellman eq.}) and solve it numerically using the value iteration algorithm to obtain the optimal policy. Further details and formal definitions are provided in Appendix \ref{Dynamic programming algo}.\\

% para3 - Explain figure 3
\begin{figure}[h]
    \centering
    \includegraphics[width=\linewidth]{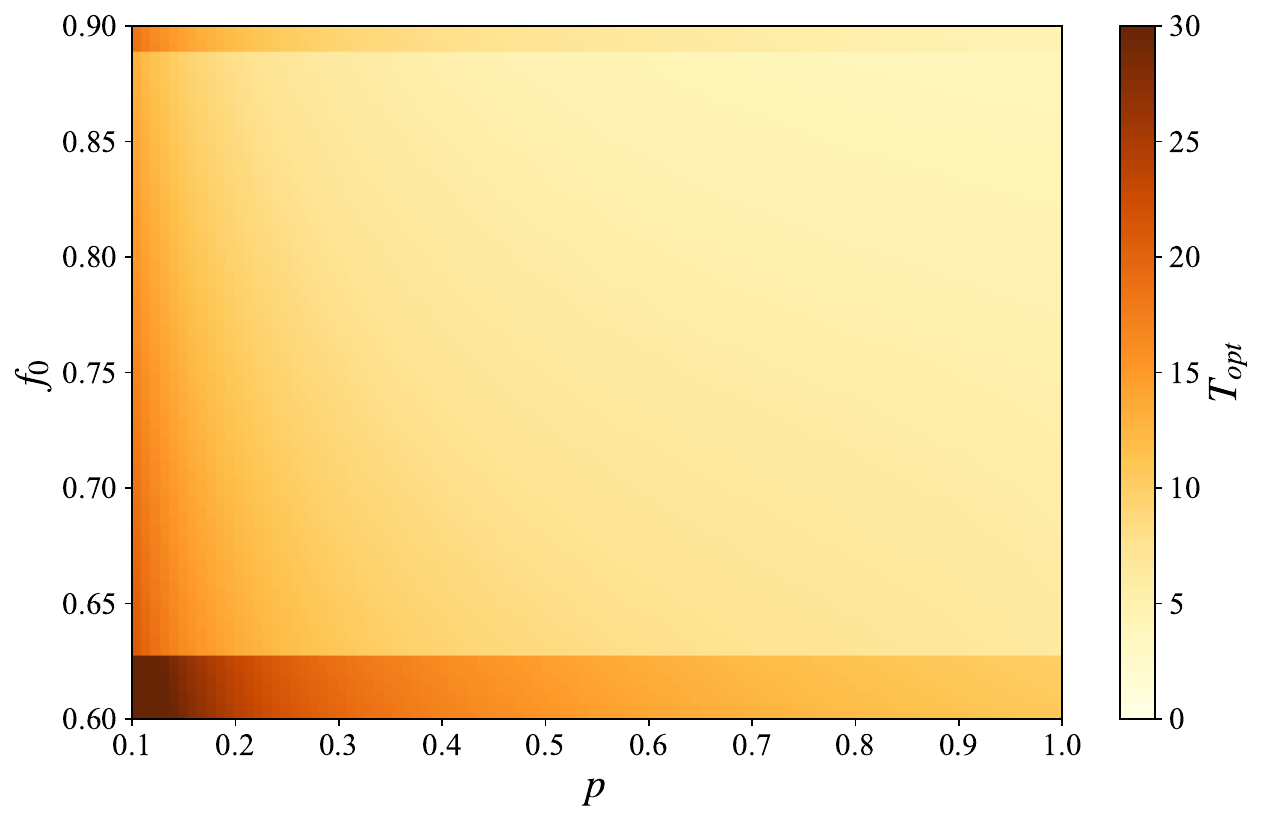}
    \caption{Expected waiting time of optimal policy, $T_{opt}$, for a node with four quantum memories ($m=4$), shown for different values of the entanglement generation probability $p$ and initial fidelity $f_0$. The gap between the target fidelity $f_T$ and the initial fidelity $f_0$ is fixed at $\Delta f = 0.04$.}
    \label{ToptVs(p,f0)}
\end{figure}

In this section, we numerically analyze the dependence of the expected waiting time under an optimal policy, denoted by $T_{opt}$, on the system parameters $(m,p, f_0,f_T)$ and compare it with baseline policies, as defined in Sec. \ref{baseline policies}. Our results show that $T_{opt}$ increases when the entanglement generation probability $p$ is small, since a larger number of attempts are required to successfully establish entangled links. For a given target fidelity $f_T$, reducing the initial fidelity $f_0$ also increases $T_{\text{opt}}$, since a larger fidelity gap must be bridged and the success probability of distillation is reduced. However, increasing the number of quantum memories $m$ reduces the $T_{\text{opt}}$, as it provides more distillation configurations to achieve the target fidelity efficiently. We further find that the advantage of the optimal policy over baseline strategies depends nontrivially on the system parameters, with its expected waiting time being comparable to that of pumping at intermediate $f_0$ and nested policy at low and high $f_0$ for a fixed fidelity gap $\Delta f$, as discussed in detail throughout this section.\\ 
% Have one plot for the Topt Vs m 

In Fig. (\ref{ToptVs(p,f0)}), we show the expected waiting time of an optimal policy for a node with four quantum memories, as a function of initial fidelity $f_0$ and generation probability $p$. For fixed fidelity gap, $\Delta f = (f_T - f_0)$, increasing $p$ reduces the optimal expected waiting time, $T_{opt}$. However, $T_{opt}$ exhibits non-monotonic behavior with respect to $f_0$. Specifically, $T_{opt}$ is higher at both low ($f_0 \sim 0.6$) and high values ($f_0 \sim 0.9$) of initial fidelity, while it attains lower values in an intermediate regime ($0.65 \lesssim f_0 \lesssim 0.85$). The observed discontinuities in $T_{opt}$ may arise from sudden changes in size of MDP state space. For a fixed $\Delta f$, the number of accessible fidelity levels, and hence the size of the state space is smaller in the intermediate regime of $f_0$ compared to the boundary values, leading to reduced expected waiting times. Although the state space size may be similar at the boundary values, $T_{opt}$ is larger at low $f_0$ due to lower success probability of distillation. We conjecture that increasing $\Delta f$ may lead to more gradual variation in the state space and consequently smoother behavior of $T_{opt}$. However, such regimes are computationally challenging to explore using the value iteration algorithm due to the rapid growth of state space, see Appendix \ref{Dynamic programming algo}.\\

% para4 - introduce baseline policy, explain how optimal policy is better through figure 1.
To evaluate the performance gain achievable with the optimal policy, we adopt several heuristic distillation strategies as baseline policies for benchmarking. In this work, we consider some commonly used baseline policies such as greedy, pumping, and nested policy, as described in Sec. \ref{baseline policies}. These policies are easy to implement with simple operational rules that perform well only in specific parameter regimes. For instance, the greedy policy prioritizes immediate distillation whenever at least two entangled pairs are available, which may reduce the waiting time by freeing quantum memories early and enabling additional entanglement generation attempts. However, this strategy may also be suboptimal, as premature distillation can involve pairs with mismatched fidelities, leading to lower success probabilities and therefore inefficient use of resources. Similarly, pumping and nested purification perform well in memory-constrained and high-memory, parallel-operation regimes, respectively, but are not universally optimal. Given these trade-offs, it is nontrivial to determine the conditions under which the performance of such heuristic policies approximate the optimal policy. We therefore compare the expected waiting times of optimal policy with these baseline policies.\\       
% para5 - Explain fig 4
\begin{figure}[h]
    \centering
    \includegraphics[width=\linewidth]{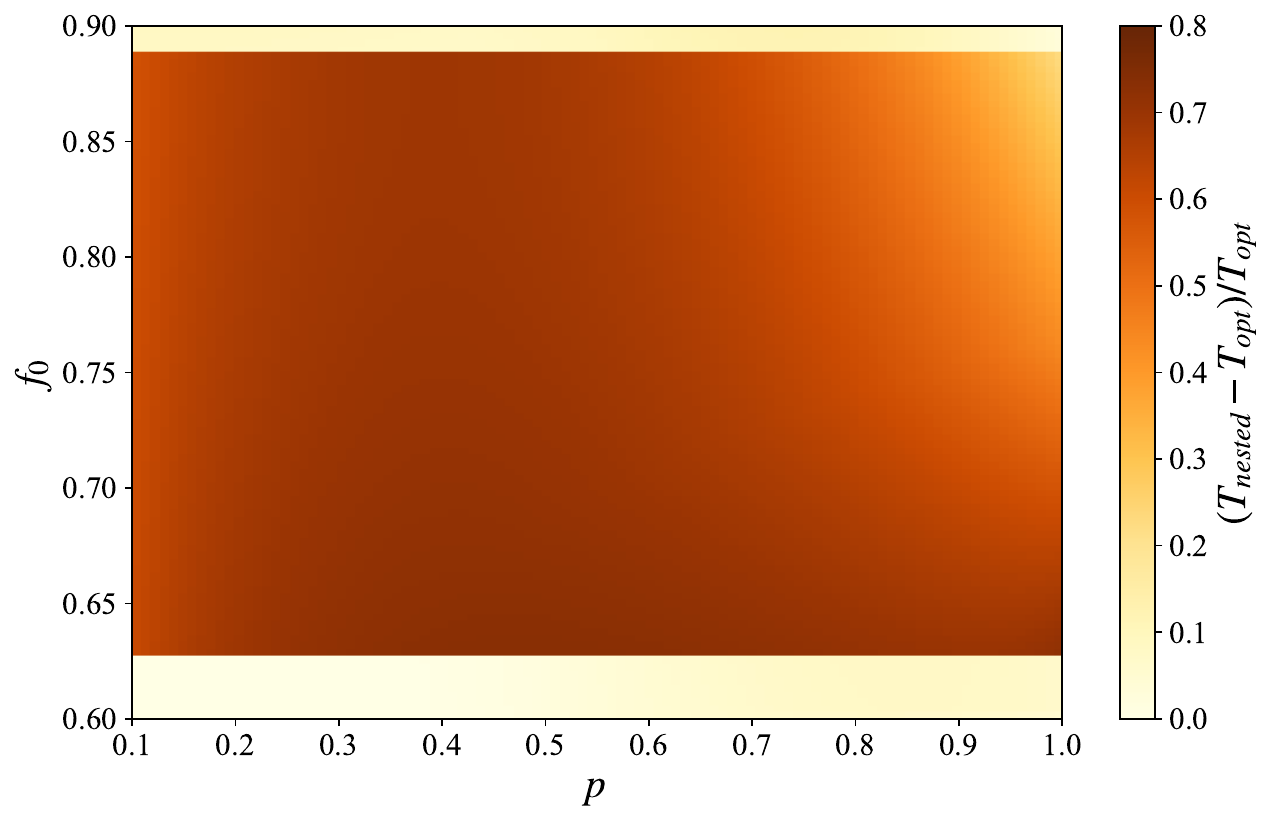}
    \caption{The relative difference between the expected waiting time of optimal policy, $T_{\text{opt}}$, and nested policy, $T_{\text{nested}}$, with four quantum memories per node ($m=4$), for different values of $p$ and $f_0$. For a fixed fidelity gap between initial and target fidelity ($\Delta f = 0.04 $), an optimal policy performs increasingly better than nested policy in regimes of intermediate $f_0$.}
    \label{nested}
\end{figure}

Fig. (\ref{nested}) illustrates the relative difference between the expected waiting times of the optimal policy, $T_{opt}$, and the nested policy, $T_{nested}$, for a node with four quantum memories and fixed fidelity gap $\Delta f$. The quantity $(T_{\text{nested}} - T_{\text{opt}})/T_{\text{opt}}$ quantifies the performance advantage of optimal policy, with larger values indicating a greater reduction in expected waiting time. Our results show that, under multiple parallel 2-to-1 ED per step and fixed $\Delta f$, the optimal policy can reduce the expected waiting time by up to $80 \%$ relative to the nested policy at intermediate values of $f_0$, across all generation probabilities $p$. For intermediate values of $f_0$, the smaller number of accessible fidelity levels reduces the availability of compatible same-fidelity pairs, causing the nested policy to spend longer waiting times for distillation opportunities. The optimal policy avoids this restriction by selecting distillation operations based on the full memory configuration. In contrast, at low and high values of $f_0$, compatible pairs occur more frequently due to the availability of larger state space, allowing the nested policy to achieve performance closer to optimal. Beyond fixed $\Delta f$, as the fidelity gap $\Delta f$ is varied, the advantage of the optimal policy over nested policy depends sensitively on the system parameters ($p,f_0,m$). For instance, at $f_0 = 0.75$ and $f_T < f_{\infty}(f_0)$, the advantage varies non-monotonically with $\Delta f$ for all values of $p$.\\

% para5 - Explain fig 5
\begin{figure}[h]
    \centering
    \includegraphics[width=\linewidth]{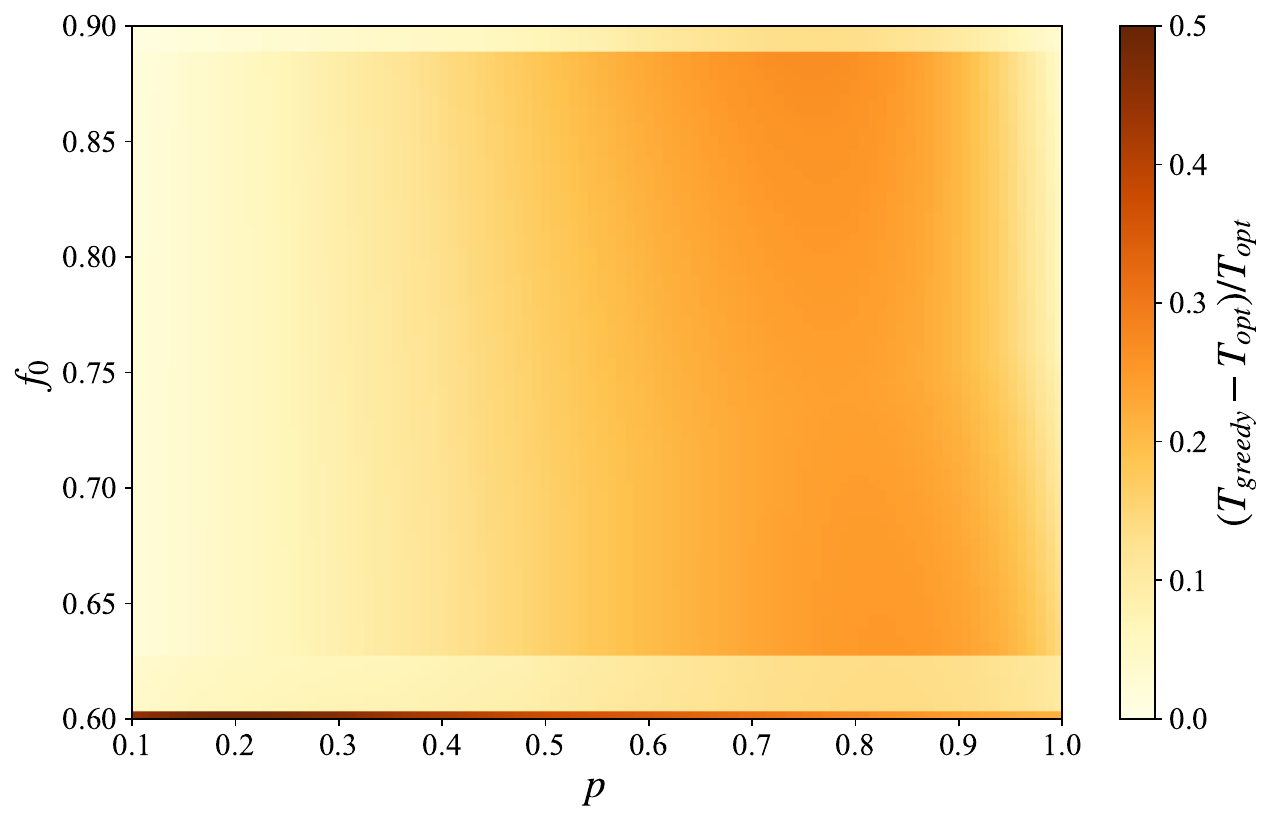}
    \caption{The relative difference between the expected waiting time of optimal policy, $T_{\text{opt}}$, and greedy policy, $T_{\text{greedy}}$, with four quantum memories per node ($m=4$), for different values of $p$ and $f_0$. For fixed fidelity gap between initial and target fidelity ($\Delta f = 0.04 $), an optimal policy performs increasingly better than greedy policy in regimes of intermediate $f_0$, as well as low $p$ with very low $f_0$.}
    \label{greedy}
\end{figure}

Fig. (\ref{greedy}) shows the relative difference between the expected waiting times of the optimal policy, $T_{opt}$, and the greedy policy, $T_{greedy}$, for a node with four quantum memories and fixed fidelity gap $\Delta f$.
As discussed in Sec. \ref{baseline policies}, the greedy policy may be viewed as a relaxed version of the nested policy, performing distillation between the pair of active links having the smallest fidelity difference whenever possible.
Although less restrictive than nested purification, the greedy policy remains myopic and may perform distillation operations that are suboptimal from a long-term perspective. As a result, the performance advantage of the optimal policy exhibits a trend similar to that observed for nested purification, but with a smaller magnitude due to the less restrictive matching criterion of the greedy policy. At intermediate values of $f_0$, the optimal policy achieves up to a $30\%$ reduction in expected waiting time compared with the greedy policy, as the smaller number of accessible fidelity levels makes it more difficult for the greedy strategy to find well-matched distillation pairs. However, at low and high values of $f_0$, compatible pairs are more readily available, allowing the greedy policy to achieve performance closer to optimal. A narrow strip-like region of enhanced advantage appears at very low $f_0$ and low p, which may be attributed to the low success probability of distillation in this regime. As a result, the greedy policy incurs frequent regeneration cycles, while the optimal policy may benefit from more selective action choices. \\

% para5 - Explain fig 5
\begin{figure}[h]
    \centering
    \includegraphics[width=\linewidth]{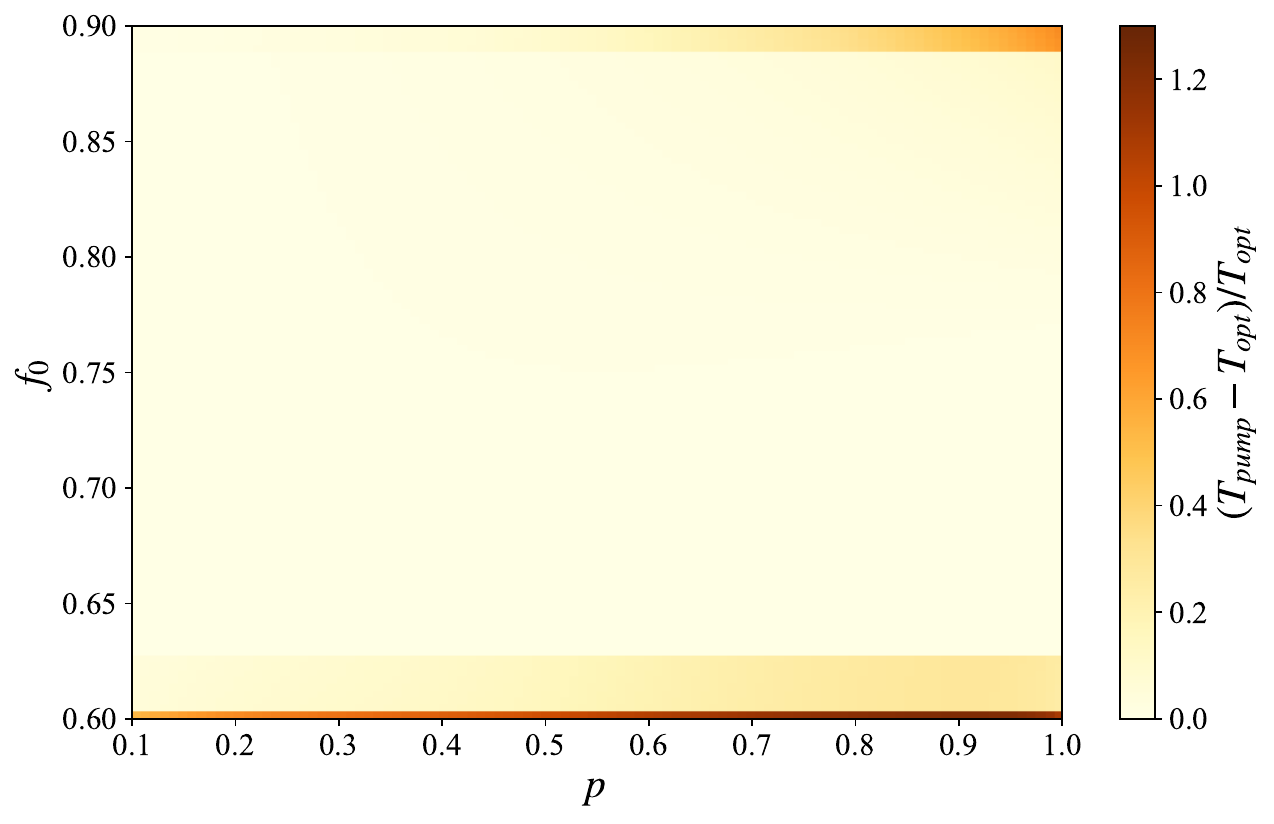}
    \caption{The relative difference between the expected waiting time of optimal policy, $T_{\text{opt}}$, and pumping policy, $T_{\text{pump}}$, with four quantum memories per node ($m=4$), for different values of $p$ and $f_0$. For fixed fidelity gap between initial and target fidelity ($\Delta f = 0.04 $), an optimal policy performs increasingly better than pumping policy in regimes of high $p$ with low and high $f_0$.}
    \label{pump}
\end{figure}

The relative difference between the expected waiting times of the optimal policy, $T_{opt}$, and the pumping policy, $T_{pump}$, is shown in Fig. (\ref{pump}), for a node with four quantum memories at fixed $\Delta f$. 
At intermediate values of $f_0$, the two policies achieve comparable expected waiting times. In this regime, the number of accessible fidelity levels is relatively small, limiting the available distillation configurations and causing the optimal policy to adopt a strategy similar to pumping. In contrast, the advantage of the optimal policy becomes more pronounced at low and high values of $f_0$. 
As discussed earlier, the optimal policy tends to behave similarly to the nested policy in these regimes, making it more effective than the pumping strategy. In this regime, the rigid sequential structure of pumping fails to fully exploit the available memory resources and rapidly generated entangled links. Consequently, the optimal policy achieves lower expected waiting times, with the advantage increasing at higher p as pumping repeatedly distills newly generated $f_0$. Upon varying the fidelity gap $\Delta f$, the advantage of the optimal policy over the pumping policy also depends sensitively on the system parameters ($p,m,f_0$). In particular, for $f_0 = 0.75$ and $f_T < f_{\infty}(f_0) $, the advantage increases with the fidelity gap across all $p$.\\

\begin{figure}[h]
    \centering
    \includegraphics[width=\linewidth]{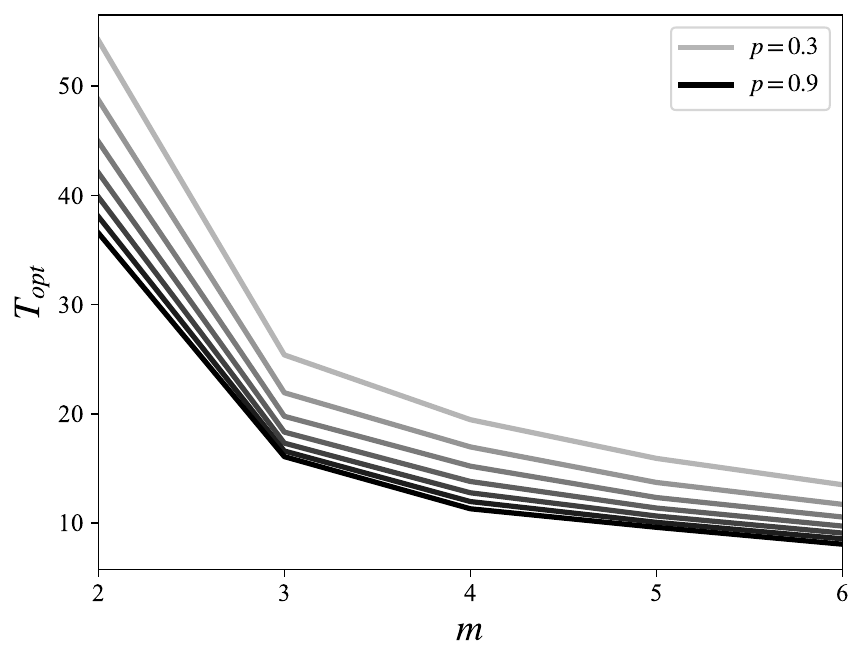}
    \caption{The behaviour of expected waiting time of optimal policy, $T_{opt}$, with number of quantum memories $m$, for different values of entanglement generation probabilities $p$. Black lines correspond to $p = 0.9$, and the value of p decreases in steps of 0.1 with increasing line transparency up to $p = 0.3$. The initial fidelity $f_0$ is 0.6 with fidelity gap of $0.04$.}
    \label{ToptVsm}
\end{figure}

Fig. (\ref{ToptVsm}) shows the expected waiting time of the optimal policy $T_{opt}$, with the number of memories $m$ for different generation probabilities, $p$. The expected waiting time, $T_{opt}$, exhibits a monotonic decrease with increasing $m$. 
 For fixed $p$, larger memory ensemble expands the state space and provides more simultaneous opportunities for entanglement generation and parallel distillation, enabling the system to reach the target fidelity faster. For a fixed $m$, $T_{opt}$ decreases with increasing $p$, since higher generation probability allows active links to be replenished more quickly, minimizing idle waiting times and enabling fidelity to build up faster.

%%%%%%%%%%%%%%%%%%%%%%%%%%%%%%%%%%%%%

%%%%%%%%%%%%%%%%%%%%%%%%%%%%%%%%%%%%%
\section{Conclusions and Future Work}
\label{conclusions}

In this work, we studied the problem of optimizing entanglement distillation policies between two remote parties, each equipped with multiple quantum memories connected by parallel quantum channels. By formulating this task as a Markov decision process (MDP), we provided a systematic framework to identify optimal entanglement distillation strategies using value iteration algorithm. The resulting optimal deterministic policy minimizes the expected waiting time required to reach the target fidelity $f_T$ based on the current system configuration.

In our analysis, we adopted a simplified yet physically motivated model, i.e., entanglement generation is assumed to stochastically produce Werner states, obtained via a depolarizing channel acting on maximally entangled Bell states and for entanglement distillation, we consider a 2-to-1 ED protocol \cite{Deutsch1996DEJMPS}. Within this framework, parallel entanglement generation can be performed simultaneously across inactive links, while multiple 2-to-1 distillation operations are allowed in parallel on disjoint pairs of active links. In certain hardware architectures, practical limitations such as crosstalk and correlated errors \cite{Sarovar2020,Heinz2021,Cheng2024} may restrict the system to at most a single distillation operation per decision step. Such constrained settings can also be naturally incorporated within our MDP framework, which identifies the optimal policy under the imposed operational constraints.

Our numerical results demonstrate that the expected waiting time under the optimal policy decreases with increasing entanglement generation probability $p$ and the number of quantum memories $m$, as both enhance the availability of entangled resources and distillation opportunities. In contrast, the expected waiting time exhibits a non-monotonic dependence on the initial fidelity $f_0$ for fixed $\Delta f$. We further show that the optimal policy can substantially reduce the expected waiting time compared with heuristic baseline strategies, including entanglement pumping, nested and greedy policy, with the magnitude of the improvement depending on the system parameters $(m,p,f_0,f_T)$. For a fixed fidelity gap ($\Delta f \sim 0.04$), the expected waiting time under the optimal policy is comparable to that of pumping for intermediate initial fidelities ($0.65 \lesssim f_0 \lesssim 0.85$), while it approaches the performance of nested purification at low ($f_0 \sim 0.6$) and high ($f_0 \sim 0.9$) fidelities. However, the variation of the optimal policy's advantage in expected waiting time with the fidelity gap $\Delta f$ depends nontrivially on both the chosen baseline policy and the system parameters ($p,m,f_0$). These findings highlight the importance of carefully balancing the trade-off between immediate distillation and delayed operations, a feature that is not captured by simple heuristic policies.

%Our numerical results demonstrate that optimal policies can significantly outperform heuristic  baseline strategies, including entanglement pumping, nested purification, greedy, and banded purification schemes. In particular, we observe that the advantage of optimal policy becomes increasingly important in regimes with higher probability of successful  entanglement generation, higher initial fidelity, and larger numbers of quantum memories. These findings shows the importance of carefully balancing the trade-off between immediate distillation and delayed operations, which is not captured by simple heuristic policy.

Despite these advantages, our approach has certain limitations. The value iteration algorithm is applicable only within a restricted regime where the state space remains finite, which in our formulation requires that the target fidelity is below the asymptotic fidelity $f_{\infty}(f_0)$ \cite{RSS}, as described in Eq. (\ref{finf}). This limitation can be partially addressed by discretizing the fidelity gap into small bins, where each bin represents a state in the MDP, thereby providing an approximate estimate of the expected waiting time. However, the size of the state space grows exponentially with the number of quantum memories, scaling as $|S| \sim (K(f_0,f_T)+2)^m$, where $K(f_0,f_T)$ is the non-linear function of $f_0$ and $f_T$. As a result, exact dynamic programming methods become computationally intractable for larger memory sizes. In such cases, where the state space is large but the action space remains finite, approximate solution methods based on function approximation, such as deep Q-networks (DQN), provide a promising alternative for learning near-optimal policies at a significantly reduced computational cost \cite{Mnih2015}.

An interesting extension of this work would be to incorporate memory decoherence into the framework, thereby providing a more realistic description of quantum hardware. In addition, while the present model considers only two end nodes, it is natural to generalize this setting to multi-node quantum repeaters, where intermediate nodes assist in entanglement distribution between Alice and Bob. Furthermore, an important direction is to account for operational imperfections such as measurement errors. In this case, the problem of optimizing entanglement distillation policy can be formulated as a partially observable Markov decision process (POMDP), which naturally captures uncertainty regarding the system state and provides a more realistic decision-making framework.

%%%%%%%%%%%%%%%%%%%%%%%%%%%%%%%%%%%%%
%%%%%%%%%%%%%%%%%%%%%%%%%%%%%%%%%%%%%
\section*{Acknowledgments}
Funding from DST, Government of India through the
SERB Grant No. MTR/2022/000389, IITB TRUST Labs
Grant No. DO/2023-SBST002-007, and the IITB seed funding is gratefully acknowledged.

%%%%%%%%%%%%%%%%%%%%%%%%%%%%%%%%%%%%%
%%%%%%%%%%%%%%%%%%%%%%%%%%%%%%%%%%%%%
\section*{Data Availability}
The data are available from the authors upon reasonable request.

%%%%%%%%%%%%%%%%%%%%%%%%%%%%%%%%%%%%%
%%%%%%%%%%%%%%%%%%%%%%%%%%%%%%%%%%%%%

\appendix

\section{Dynamic programming algorithm}
\label{Dynamic programming algo}

To determine the optimal ED policy, we formulate the problem as a MDP, which leads to Bellman optimality equations, as described in the main text. These equations can be solved using a dynamic programming algorithm, such as value iteration. The algorithm starts with arbitrary values of $T_{\pi}(s)$, for some policy $\pi$ and $\forall s \in \mathcal{S},$ where $\mathcal{S}$ is the state space and then iteratively updates the policy $\pi$ and value $T_{\pi}(s)$, $\forall s\in S$. The updated policy is guaranteed to converge to an optimal policy $\pi^{*}$ for finite MDP, see \cite{SuttonBarto2018RL}. There may exist multiple optimal policies that achieve the same minimum expected waiting time where, value iteration returns one such near-optimal policy. In practice, the iterations are terminated once the change in the expected time between successive iterations falls below a prescribed tolerance value, $\epsilon > 0$, as seen in the Algorithm $1$. In all our numerical calculations, we use $\epsilon = 10^{-6}$.\\

%Our implementation of optimal policy can be found at \textcolor{blue}{repo}.\\

\begin{algorithm}[tbh]
\centering
\rule{\linewidth}{0.8pt}

\vspace{-7pt}

\makeatletter
\renewcommand{\fnum@algorithm}{\textbf{Algorithm \thealgorithm}}
\makeatother

\caption{Value Iteration Algorithm}
\vspace{-7pt}

\rule{\linewidth}{0.4pt}
\vspace{-5pt}
\begin{algorithmic}[1]

\State Initialize $T_0(s)=0$ for all $s \in \mathcal{S}$ and small $\epsilon$

\Repeat
    \State $\delta \gets 0$
    \For{all non-terminal states $s \in \mathcal{S}$}
        \State 
        \[
        T_{k+1}(s) \gets \max_{a} \big[r(s,a) + \gamma \sum_{s',r} p(s',r \mid s,a) T_k(s') \big]
        \]
        \State $ \delta \gets \max\big(\delta, |T_{k+1}(s) - T_{k}(s)|\big) $
\EndFor
\Until{$\delta < \epsilon$}

\State \textbf{Output:} Deterministic policy $\pi^*$ such that
\[
\pi^{*}(s) = \arg\max_{a} \big[ r(s,a) + \gamma\sum_{s',r} p(s',r \mid s,a) T(s') \big]
\]

\end{algorithmic}

\vspace{-5pt}
\rule{\linewidth}{0.8pt}

\label{Algorithm}
\end{algorithm}

The computational cost of the algorithm grows rapidly with the size of the state and action spaces. In general, the computational complexity of a single iteration of value iteration scales as $O(|\mathcal{A}||\mathcal{S}|^2)$, where $\mathcal{S}$ is the state space and $\mathcal{A}$ is the action space \cite{kaelbling1996reinforcement}. In our problem, the size of the state space scales as $|\mathcal{S}| = (K(f_0,f_T) + 2)^m$, where $m$ is the number of quantum memories and $K(f_0,f_T)$ is the number of distinct fidelity levels between the initial fidelity $f_0$ and the target fidelity $f_T$. The action space scales as $O(m^{\lfloor m/2 \rfloor}.2^m)$ due to the possible pairwise distillation choices. Therefore, the overall computational complexity increases exponentially with the number of memories and the number of fidelity levels. Moreover, when the entanglement generation probability $p$ is small, the effective horizon of the process increases, leading to slower convergence of the value iteration algorithm. Due to this scaling, exact dynamic programming becomes computationally intractable for systems with a large number of quantum memories, large fidelity gaps, or low entanglement generation probabilities. In such regimes, approximate methods, such as deep reinforcement learning \cite{SuttonBarto2018RL,Mnih2015}, provides sub-optimal policies at a significantly reduced computational cost.

\begin{comment}

\section{Scaling of Number of States}
\label{number of states scaling}

% para1 -  scaling of the number of states with the number of memories.

In this appendix, we analyze the scaling of the number of states in the Markov decision process discussed in the main text. The size of state space mainly depends upon number of quantum memories at each station, initial fidelity and threshold fidelity. The number of states in state space $\mathcal{S}$,  scales as follow, 

\begin{align}
    |\mathcal{S}| = (K(f_i,f_t) + 2)^m
\end{align}

As seen above that, for given initial and threshold fidelity, the number of states grows exponentially with the number of memories. 

As discussed above, we get the exact number of the states in the regime where threshold fidelity less than asymptotic fidelity of the initial fidelity. In order to pass that limitation, we developed the approximate method by dividing the fidelity gap between initial fidelity and threshold fidelity in small bins and each bin is state in the MDP. The overall scaling of state space with $f_i$ and $f_t$ is shown in figure ().

% para - Scaling of the number of states with the initial and thresold fidelity. (exact (ft<fasy) and approximate(ft>fasy))-

\end{comment}

%%%%%%%%%%%%%%%%%%%%%%%%%%%%%%%%%%%%%
%%%%%%%%%%%%%%%%%%%%%%%%%%%%%%%%%%%%%

%\bibliographystyle{apsrev4-1}
\bibliography{references.bib}
\end{document}